\documentclass[twocolumn,prb,showpacs,floatfix]{revtex4}
\usepackage{graphicx}
\usepackage{amssymb,amsmath,latexsym}
\usepackage{bm}
\usepackage[dvips]{epsfig,color}

\begin{document}

%%%%%%%%%%%%%%%%%%%%%%%%%%%%%%%%%%%%%%%%%%%%%%%%%%%%%%%%%%%%%%%%%%%%
%\title{Generic Electron-Hole Asymmetric Spin-Orbit Coupling in Carbon Nanotubes}
%\title{Generic electron-hole asymmetric spin-orbit coupling in ballistic semiconducting carbon nanotubes}
%\title{\blue{Two types of curvature-enhanced spin-orbit coupling in carbon nanotubes}}
\title{Curvature-enhanced spin-orbit coupling in a carbon nanotube}
\author{Jae-Seung Jeong and Hyun-Woo Lee}
\affiliation{PCTP and Department of Physics, Pohang University of Science and Technology, Pohang, 790-784, Korea}
%\affiliation{$^2$Department of electrical and computer engineering,
%National University of Singapore, Singapore 117576}

\date{\today}

%%%%%%%%%%%%%%%%%%%%%%%%%%%%%%%%%%%%%%%%%%%%%%%%%%%%%%%%%%%%%%%%%%%%%

\begin{abstract}
Structure of the spin-orbit coupling varies from material to
material and thus finding the correct spin-orbit coupling structure
is an important step towards advanced spintronic applications. We
show theoretically that the curvature in a carbon nanotube generates
two types of the spin-orbit coupling, one of which was not
recognized before. In addition to the topological phase-related
contribution of the spin-orbit coupling, which appears in the
off-diagonal part of the effective Dirac Hamiltonian of carbon
nanotubes, there is another contribution that appears in the
diagonal part. The existence of the diagonal term can modify
spin-orbit coupling effects qualitatively, an example of which is
the electron-hole asymmetric spin splitting observed recently, and
generate four qualitatively different behavior of energy level
dependence on parallel magnetic field. It is demonstrated that the
diagonal term applies to a curved graphene as well. This
result should be valuable for spintronic applications of graphitic
materials.
%The interplay of the atomic spin-orbit coupling and the curvature considerably enhances
%the effective spin-orbit coupling (SOC) in graphitic materials such as carbon nanotubes and graphenes.
%and their understanding is important for spintronic application of a carbon nanotube and a graphene.
%We show theoretically that the interplay gives to two types of the
%effective SOC,
%one of which was not recognized previously.
%The second type of the effective SOC is comparable in magnitude to the first type
%and modifies the spin splitting pattern in a carbon nanotube qualitatively.
%This result should be useful for spintronic applications in graphitic materials.
%
%The curvature and topology of carbon nanotubes (CNTs) play important roles
%for their spin-orbit coupling (SOC). While existing
%theories predict the same coupling strengths for electrons and
%holes, a recent experiment
%%[F. Kuemmeth, S. Ilani, D. C. Ralph $\&$ P. L. McEuen, Nature {\bf 452}, 448 (2008)]
%found significant difference. Here we show theoretically that in
%addition to the topological phase-related contribution of the
%SOC, which appears in the off-diagonal part of the
%effective Dirac Hamiltonian of CNTs, there is another
%contribution that appears in the diagonal part. The interplay
%between the two contributions induces the electron-hole asymmetry
%in the SOC strength and generates four different behavior of energy level
%dependence on parallel magnetic field.
%Our theory
%explains the recent experiment and should be useful for spintronic
%applications of CNTs.
\end{abstract}

%\pacs{}

\maketitle
%%%%%%%%%%%%%%%%%%%%%%%%%%%%%%%%%%%%%%%%%%%%%%%%%%%%%%%%%%%%%%%%%%%%%%%%%%

%{\it Introduction.--}

\section{introduction}
Graphitic materials such as carbon nanotubes (CNTs) and graphenes are promising materials for spintronic applications.
Various types of spintronic devices are reported such as
CNT-based three terminal magnetic tunnel junctions~\cite{Sahoo05NaturePhysics}, spin diodes~\cite{Merchant08PRL}, and
graphene-based spin valves~\cite{Tombros07Nature}.
Graphitic materials are believed to be excellent spin conductors~\cite{Tsukagoshi99Nature}.
%with long spin lifetime.
The hyperfine interaction of electron spins with nuclear spins is strongly suppressed
since $^{12}$C atoms do not carry nuclear spins.
It is estimated that the spin relaxation time in a CNT~\cite{Bulaev08PRB}
and a graphene~\cite{Hernando08Condmat}
%of graphitic materials
is limited by the spin-orbit coupling (SOC).
%\comment{Here we need to cite a paper that analyzes the effect of the SOC on the spin relaxation in CNT.}

Carbon atoms are subject to the atomic SOC Hamiltonian $H_{\rm so}$.
%Carbon atoms are subject to the atomic SOC Hamiltonian $H_{\rm so}\!=\!
%\lambda_{\rm so}\sum_r \mathbf{L}_{r} \cdot \mathbf{S}_{r}$, where
%$\mathbf{L}_{r}$ and $\mathbf{S}_{r}$ are respectively the atomic
%orbital and spin angular momentum of an electron at a carbon atom
%$r$, and $\lambda_{\rm so}\!\sim\!10\,$meV~\cite{Serrano00SSC} is the atomic SOC constant.
In an ideal flat graphene,
the energy shift caused by $H_{\rm so}$ is predicted to be $\sim 10^{-3}$ meV~\cite{Min06PRB,Hernando06PRB}.
%In flat graphitic materials such as ideal flat graphenes,
%the energy shift caused by $H_{\rm so}$ is much smaller ($\sim 10^{-3}$ meV) than $\lambda_{\rm so}$
%since $H_{\rm so}$ does not cause intra-band transition within the $\pi$ bands near the Fermi level.
Recently it is predicted~\cite{Hernando06PRB,Ando00JPSJ} that
the geometric curvature can enhance
the effective strength of the SOC by orders of magnitude.
% (for instance, $\sim\!0.1$meV for CNTs with the diameter of $\sim\!5$nm)
%For a CNT with the diameter of $\sim\!5$ nm, the enhancement factor is of the order of 100.
This mechanism applies to a CNT and also to a graphene which, in
many experimental situations,
exhibits nanometer-scale corrugations~\cite{Meyer07Nature}.
%contains significant amount of
%ripples
%which are helpful for its mechanical stability~\cite{Kim09Nature}.
There is also a suggestion~\cite{Pereira08arXiv}
that artificial curved structures of a graphene may facilitate device applications.

A recent experiment~\cite{Kuemmeth08Nature} on ultra-clean CNTs measured directly
the energy shifts caused by the SOC, which provides an ideal opportunity to
test theories of the curvature-enhanced SOC in graphitic materials.
The measured shifts are
in order-of-magnitude agreement with the theoretical predictions~\cite{Ando00JPSJ,Hernando06PRB},
confirming that the curvature indeed enhances the effective SOC strength.
The experiment revealed discrepancies as well;
While existing theories predict the same strength of the SOC for electrons and holes,
which is natural considering that both the conduction and valence bands originate from the same $\pi$ orbital,
the experiment found considerable asymmetry in the SOC strength between electrons and holes.
%surprisingly that the energy shifts are about two times larger for electrons than for holes.
This electron-hole asymmetry
%in the SOC strength
implies that existing theories of the SOC in graphitic materials are incomplete.
%Considering that the curvature-enhanced SOC plays an important role in curved graphene structures as well,
%the resolution of this discrepancy is important for graphene spintronics as well.

In this paper, we
%analyze the interplay of the atomic SOC and the curvature in a single-walled CNT, and find
show theoretically
that in addition to effective SOC in
the off-diagonal part of the effective Dirac Hamiltonian,
which was reported in the existing theories~\cite{Ando00JPSJ,Hernando06PRB},
there exists an additional type of the SOC that appears in the diagonal part
both in CNTs and curved graphenes.
%, which is comparable in magnitude to the first type.
%and that the interplay of the two SOCs produces the electron-hole asymmetric SOC in a CNT.
It is demonstrated that the combined action of the two types of the SOC produces the electron-hole asymmetry
observed in the CNT experiment~\cite{Kuemmeth08Nature}
and gives rise to four qualitatively different behavior of energy level
dependence on magnetic field parallel to the CNT axis.

This paper is organized as follows.
In Sec.~\ref{sec:Effective SOC}, we show analytical expressions of
two types of the effective SOC in a CNT and then explain
how the electron-hole asymmetric spin splitting can be generated in
semiconducting CNTs generically.
Section~\ref{sec:theory}
describes the second-order perturbation theory
that is used to calculate the effective SOC,
and tight-binding models of the atomic SOC and geometric curvature.
Section~\ref{sec:Behavior in a magnetic field} reports
four distinct energy level dependence on magnetic field parallel to the CNT axis.
We conclude in Sec.~\ref{sec:Discussion and summary} with
implications of our theory on curved graphenes and a brief summary.

%Although both types of the effective SOCs arise from the interplay between the atomic SOC ($H_{\rm so}$) and the curvature,
%their natures in terms of the pseudospin (spanned by two sublattices $A$ and $B$ of a graphene layer)
%are different, which leads to different spin splittings in the conduction and valence bands of a CNT
%since the two bands have different pseudospin characters. This explains the origin of the electron-hole asymmetry in
%the SOC strength in the experiment~\cite{Kuemmeth08Nature}.
%Below we use a single walled CNT as a specific example to demonstrate the existence of the second type of the SOC
%and illustrate combined effects of the two SOCs.
%We also discuss briefly the second type of the SOC in a graphene.
%which can be easily derived from the CNT calculation after minor modifications.

\section{Effective spin-orbit coupling in a CNT} \label{sec:Effective SOC}
We begin our discussion by presenting the first main result for a CNT
with the radius $R$
and the chiral angle $\theta$ ($0\!\le\!\theta\!\le\!\pi/6$, $0\,(\pi/6)$ for zigzag (armchair) CNTs).
We find that when the two sublattices $A$ and $B$ of the CNT are used as bases,
the curvature-enhanced effective SOC Hamiltonian ${\cal H}_{\rm soc}$
near the K point with Bloch momentum ${\bf K}$ becomes
%
%
%
%\begin{equation}
%{\cal H}_{\rm soc}^{\rm K}=( \delta {\tau_+}+\delta^* \tau_-)\frac{\sigma_y}{2R}+\delta'\tau_{\rm I}\frac{\sigma_y}{R},
%\label{eq:effective SOC Hamiltonian}
%\end{equation}
%
%
%
\begin{equation}
{\cal H}_{\rm soc}^{\rm K} = \left(  \begin{array}{cc}
(\delta'_{\rm K}/R)\sigma_y & (\delta_{\rm K}/R)\sigma_y \\
(\delta^{*}_{\rm K}/R)\sigma_y & (\delta'_{\rm K}/R)\sigma_y
\end{array} \right) ,
\label{eq:effective SOC Hamiltonian}
\end{equation}
where
$\sigma_y$ represents the real spin Pauli matrix along the CNT axis.
The pseudospin is defined to be up (down) when an electron
is in the sublattice $A\,(B)$.
Here the off-diagonal term that can be described by a spin-dependent
topological phase
are reported in Refs.~\cite{Hernando06PRB,Ando00JPSJ}
but
the diagonal term was not recognized before.
Expressions for the parameters $\delta_{\rm K}$ and $\delta'_{\rm K}$ are
given
by~\cite{comment-delta}
%
%
%%%%%%%%%%%%%%%%%%%%%%%%%%
\begin{equation}
{\delta_{\rm K}\over R} =
%\!\!\left\langle\psi_{A}^{\rm K}\right|\!H^{\rm K,(2)}\!\left|\psi_{B}^{\rm K}\right\rangle
%\!\=
\frac{\lambda_{\rm so}a(\varepsilon_{s}-\varepsilon_{p})(V_{pp}^{\pi}+V_{pp}^{\sigma})}
{12\sqrt{3}{V_{sp}^{\sigma}}^2}\frac{e^{-i\theta}}{R}
\label{eq:delta}
\end{equation}
%%%%%%%%%%%%%%%%%%%%%%%%%%%%
%
%
and
\begin{equation}
{\delta'_{\rm K}\over R} =
%\!\!\left\langle\psi_{A}^{\rm K}\right|\!H^{{\rm K},(2)}\!\left|\psi_{A}^{\rm K}\right\rangle
%\!=
\frac{\lambda_{\rm so}a V_{pp}^{\pi}}{2\sqrt{3}(V_{pp}^{\sigma}-V_{pp}^{\pi})}
\frac{\cos 3\theta}{R},
\label{eq:delta'}
\end{equation}
where $\lambda_{\rm so}\sim 12\,$meV~\cite{Serrano00SSC} is the atomic SOC constant,
$a$ is the lattice constant $2.49{\AA}$,
and $\varepsilon_{s(p)}$ is the atomic energy for the $s(p)$ orbital.
Here,
%$V_{sp}^{\sigma}$ and $V_{pp}^{\pi(\sigma)}$ are tight-binding (TB) parameters
%explained fully in Sec.~\ref{sec:TB}}
$V_{sp}^{\sigma}$ and $V_{pp}^{\pi(\sigma)}$ represent the coupling strengths
in the absence of the curvature for the $\sigma$ coupling
between nearest neighbor $s$ and $p$ orbitals and the $\pi$($\sigma$) coupling
between nearest neighbor $p$ orbitals, respectively.
Note that the $|\delta'_{\rm K}|$ has the $\theta$-dependence,
whose implication on the CNT energy spectrum
is addressed in Sec.~\ref{sec:Behavior in a magnetic field}.
For K$'$ point with ${\bf K'}\!=\!-{\bf K}$,
${\cal H}_{\rm soc}^{\rm K'}$
is given by Eq.~(\ref{eq:effective SOC Hamiltonian})
with $\delta_{\rm K}$ and $\delta_{\rm K}'$ replaced by
$\delta_{\rm K'}\!=\!-\delta^*_{\rm K}$ and $\delta'_{\rm K'}\!=\!-\delta'_{\rm K}$,
respectively.
%The signs of $\delta e^{i\theta}$ and $\delta'\!/\!\cos3\theta$ are negative.
%We find $|(\delta'\!/\!\cos 3\theta)/\delta|\!=\!4.5$
%for tight-binding (TB) parameters in Ref.~\cite{Mintmire95Carbon}.
%\comment{Jae-Seung, please provide numerical values of this ratio}
%Thus $\delta'$ is of the same order as
%$\delta$~\cite{parameters}, which is understandable since pseudospin flipping terms
%in $E^{(0)}\!-\!H^{\rm{K},(0)}$ (with amplitudes $V_{pp}^{\sigma}$, $V_{sp}^{\sigma}$)
%are comparable in magnitude to pseudospin conserving terms
%(with amplitudes $E^{(0)}\!-\!\varepsilon_{s (p)}$).
%, and
%$\delta$, $\delta'$ are the amplitudes of the first and second types of the effective SOCs.
%Expressions for $\delta$ and $\delta'$
%are given in Eqs.~(\ref{eq:delta}) and~(\ref{eq:delta'}).
%

%%%%%%%%%%%%%%%%%%%%%%%%%%%%%%%%%%%%%%%%%%%%%%%%
\begin{figure}[b!]
\centerline{\includegraphics[width=8.5cm]{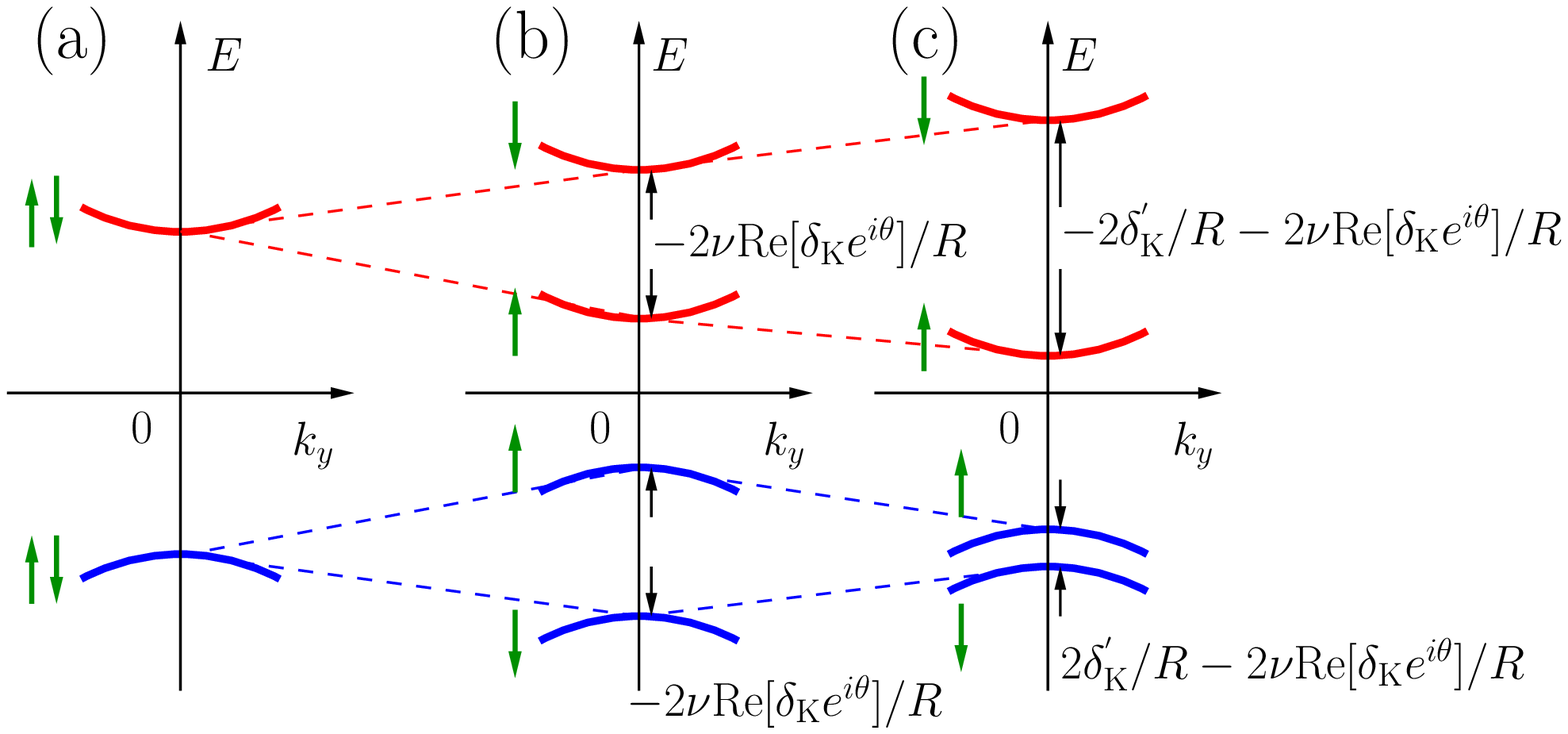}}
\caption{(Color online) Schematic diagram of the lowest conduction (red, $E>0$)
and highest valence (blue, $E<0$) band positions of a
semiconducting CNT predicted by ${\cal H}_{\rm Dirac}^{\rm
K}+{\cal H}_{\rm soc}^{\rm K}$ for (a)
$\delta_{\rm K}\!=\!\delta'_{\rm K}\!=\!0$, (b) $\delta_{\rm K}\!\neq\!0$,
$\delta'_{\rm K}\!=\!0$, and (c) $\delta_{\rm K}\!\neq\!0$, $\delta'_{\rm K}\!\neq\! 0$.
In (c), the conduction or valence band has larger spin
splitting depending on the sign of $\nu$.
Arrows (green) show the spin direction along the CNT.
%\blue{
The expressions for the energy level spacing
%at $k_y\!=\!0$
are also
provided. When they are negative, the positions of the two
spin-split bands should be swapped.
%Note that the effective SOC
%strengths for the conduction and valence bands become different when
%both $\delta$ and $\delta'$ are nonzero.
%}
%\comment{Two signs mistakes in the hole part of (c)}
%
%In (c), the positions of the two spin-split bands with smaller splitting can be swapped
%according to size of $\delta'$ and $\nu{\rm Re}\left[\delta e^{i\theta}\right]$.
%
%\red{When these numbers are negative, the positions of the two spin-split bands should be swapped in the diagram.}
%\comment {How about showing just $\nu=1$ case ?}
%\comment{Please mark as follows in the figure: (b) $-2\nu \delta \cos \theta /L$ for electron and hole.
%(c) $-2\nu \delta \cos \theta /L - 2\delta'/L$ for electron and $-2\nu \delta \cos\theta /L + 2\delta'/L$ for holes.
%In (c), please swap the position of the two spin-splitted bands in hole sector
}
\label{fig:schematic level diagram}
\end{figure}
%%%%%%%%%%%%%%%%%%%%%%%%%%%%%%%%%%%%%%%%%%%%%%%%%

Implications of the
diagonal term of the SOC
become evident
when Eq.~(\ref{eq:effective SOC Hamiltonian}) is combined
with the two-dimensional Dirac Hamiltonian ${\cal H}_{\rm Dirac}$ of the CNT.
For a state near the K point with the Bloch momentum
$\mathbf{K}+\mathbf{k}$ [$\mathbf{k}=(k_x,k_y)$, $|\mathbf{k}|\ll|\mathbf{K}|$],
${\cal H}_{\rm Dirac}$ becomes~\cite{Ajiki93JPSJ}
%
%
%%%%%%%%%%%%%%%
%\begin{equation}
%{\cal H}_{\rm Dirac}^{\rm K}\!=\!\hbar v_F \left[ e^{-i\theta}(k_x\!-\!ik_y) {\tau_+ \over 2}
%\!+\! e^{i\theta}(k_x\!+\!ik_y){\tau_- \over 2} \right],
%\end{equation}
%%%%%%%%%%%%%%%
%
%
%%%%%%%%%%%%%%%
\begin{equation}
\label{eq:DiracHamiltonian}
{\cal H}_{\rm Dirac}^{\rm K}=\hbar v_{\rm F}\left( \begin{array}{cc}
0 & e^{-i\theta}(k_x -i k_y) \\
e^{+i\theta}(k_x + i k_y) & 0
\end{array} \right),
\end{equation}
%%%%%%%%%%%%%%%
%
%
where $v_{F}$ is the Fermi velocity
and the momentum component $k_x$ along the circumference direction satisfies the quantization condition
$k_x={(1/3R})\nu$
for a $(n,m)$ CNT with $n-m=3q+\nu$ ($q\in \mathbb{Z}$ and $\nu=\pm1,0$)
and $\theta=\tan^{-1}[{\sqrt{3}m/(2n+m)}]$.
For a semiconducting $(\nu\!=\!\pm 1)$ CNT,
the diagonalization of ${\cal H}_{\rm Dirac}^{\rm K}\!+\!{\cal H}_{\rm soc}^{\rm K}$
results in different spin splittings [Fig.~\ref{fig:schematic level diagram}(c)]
of $-2\delta'_{\rm K}/R-2\nu {\rm Re}[\delta_{\rm K} e^{i\theta}]/R$
and $2\delta'_{\rm K}/R-2\nu {\rm Re}[\delta_{\rm K} e^{i\theta}]/R$
for the conduction and valence bands, respectively.
This explains the electron-hole asymmetry observed in the recent experiment~\cite{Kuemmeth08Nature}.
Here we remark that neither the off-diagonal ($\delta_{\rm K}$)
nor the diagonal ($\delta'_{\rm K}$) term of the SOC alone
can generate the electron-hole asymmetry since the two spin splittings can
differ by sign at best, which actually implies the same magnitude of the spin splitting
(see Fig.~\ref{fig:schematic level diagram} for the sign convention).
Thus the interplay of the two types is crucial for the asymmetry.

%%%%%%%%%%%%%%%%%%%%%%%%%%%%%%%%%%%%%%%%%%%%%%%%%%%%%%%%%%%%%%%%%%%%%%%%%
%\begin{figure}[b!]
%\centerline{\includegraphics[width=5cm]{graphene[1].eps}}
%\caption{{\bf [under construction]}(Color) Graphene honeycomb sheet
%with primitive lattice vectors ${\bf a_1}$ and ${\bf a_2}$.
%$\mathcal{C}_h=n\,{\bf a_1}+m\,{\bf a_2}\equiv(n,m)$ is the chiral vector,
%and $\theta$ is the chiral angle.
%The unit cell including two unit atoms $A$ and $B$ is shown as the dotted rhombus
%and the first Brillouin zone is depicted in the inset.
%$\equiv \tan^{-1}\left(\sqrt{3}m/(2n+m)\right) (0\leq |\theta|\leq \pi/6)$.
%$(x',y')$ coordinates are fixed onto the graphene and $(x,y)$ are determined in such a way that
%$x$ is along the circumferential direction and $y$ is along the NT axis.
%The coordinates for a carbon nanotube is illustrated on the right.}
%\label{graphene}
%\end{figure}
%%%%%%%%%%%%%%%%%%%%%%%%%%%%%%%%%%%%%%%%%%%%%%%%%%%%%%%%%%%%%%%%%%%%%%%%%

\section{Theory and Model}\label{sec:theory}
%
%
%\subsection{Second-order perturbation}
We calculate the $\delta_{\rm K}$ and $\delta'_{\rm K}$ analytically
using degenerate second-order perturbation theory
and treating atomic SOC and geometric curvature as perturbation.
For simplicity, we evaluate $\delta_{\rm K}$ and $\delta'_{\rm K}$ in the limit ${\bf k}\!=\!0$.
Although this limit is not strictly valid since ${\bf k}\!=\!0$
does not generally satisfy
the quantization condition on $k_x$,
one may still take this limit
since the dependence of $\delta_{\rm K}$ and $\delta'_{\rm K}$ on ${\bf k}$ is weak.
An electron at the K point is described by the total Hamiltonian
$H^{{\rm K},(0)}+H_{\rm so}+H_{\rm c}$,
where $H_{\rm c}$ describes the curvature effects and $H^{{\rm K},(0)}$ describes
the $\pi$ and $\sigma$ bands
in the absence of both $H_{\rm so}$ and $H_{\rm c}$.
The $\pi$ band eigenstates of $H^{{\rm K},(0)}$ are given by

\begin{equation}
\label{eq:pi-eigenstate}
|\Psi_{\uparrow(\downarrow)}^{{\rm K},(0)}\rangle
=\frac{1}{\sqrt{2}}\left(\nu\,e^{-i\theta}\left|\psi_{A}^{\rm K}\right\rangle
\pm \left|\psi_{B}^{\rm K}\right\rangle\right)\chi_{\uparrow(\downarrow)}
%E^{{\rm K},(0)}_{\uparrow(\downarrow)}&=&E^{(0)}=0,
\end{equation}
with the corresponding eigenvalues $E^{{\rm K},(0)}_{\uparrow(\downarrow)}\!\equiv\! E^{(0)}\!=\!0$.
Here $|\Psi_{\uparrow(\downarrow)}^{{\rm K},(0)}\rangle$ with the upper (lower) sign
amounts to the ${\bf k}=0$ limit of the eigenstate at the the conduction band bottom (valence band top).
$\chi_{\uparrow(\downarrow)}$ denotes the eigenspinor of $\sigma_y$.
$|\psi_{A(B)}^{{\rm K}}\rangle ={1\over\sqrt{N}}\sum_{r={\bf r}_{A(B)}}
e^{i{\bf K}\cdot r}\left|p_{z}^r\right\rangle$ is the orbital projection of
$|\Psi_{\uparrow(\downarrow)}^{{\rm K},(0)}\rangle$
into the sublattice $A\,(B)$,
$|p_{z}^r\rangle$ represents the $p_z$ orbital at the atomic
position $r$, and the $z$ axis is perpendicular to the CNT surface.

\begin{figure}[t!]
\centerline{\includegraphics[width=8.5cm]{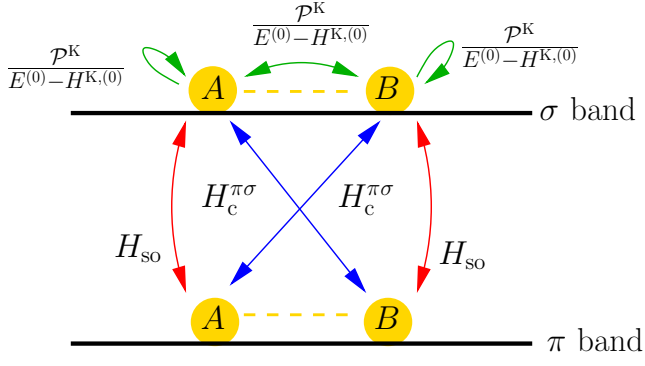}}
\caption{
(Color online)
Schematic diagram of the second order
transition process generated by $H^{{\rm K},(2)}$ [Eq.~(\ref{eq:H^K(2)})].
Pseudospin transitions (between the sublattices $A$ and $B$)
and interband transitions (between $\pi$ and $\sigma$ bands) are illustrated.
} \label{fig:transition process}
\end{figure}

When $H_{\rm so}$ and $H_{\rm c}$ are treated as weak perturbations,
the first order contribution $H_{\rm so}$ to the effective SOC vanishes
since it causes the inter-band transition (Fig.~\ref{fig:transition process}) to the $\sigma$ band~\cite{Hernando06PRB}.
The next leading order contribution to the effective SOC comes
from the following second order perturbation Hamiltonian $H^{\rm K,(2)}$~\cite{Schiff68Book},
\begin{equation}
H^{{\rm K},(2)}=H_{\rm c}\frac{\mathcal{P}^{\rm K}}{E^{(0)}
%_{ {\mathbf k}\uparrow(\downarrow) }
-H^{{\rm K},(0)}}H_{\rm so} + {\rm H.c.},
\label{eq:H^K(2)}
\end{equation}
where
%$E^{K,(0)}_{ {\mathbf k}}$ is the spin-degenerate eigenvalue of $H^{K,(0)}$ associated with $|\Psi^{K,(0)}_{ {\mathbf k}\uparrow(\downarrow)}\rangle$,
%and
the projection operator $\mathcal{P}^{\rm K}$ is defined by
$\mathcal{P}^{\rm K}\!\equiv\! 1 \!-\!\sum_{\alpha=\uparrow,\downarrow}|\Psi^{{\rm K},(0)}_{\alpha} \rangle \langle \Psi^{{\rm K},(0)}_{\alpha} |$.
Another spin-dependent second order term
$H_{\rm so}[\mathcal{P}^{\rm K}/(E^{(0)}\!-\!H^{{\rm K},(0)})]H_{\rm so}$~\cite{Second order Hamiltonian}
%However this term
is smaller than Eq.~(\ref{eq:H^K(2)}) (by two orders of magnitude for
a CNT with $R\!\sim\!2.5\,$nm), and thus ignored.
Then the second order energy shift
$E^{{\rm K},(2)}_{\uparrow(\downarrow)}$
is given by~\cite{second-order energy correction-K'}
%
%
%%%%%%%%%%%%%%%%%%
\begin{eqnarray}
\label{eq:second-order energy correction[2]}
E^{{\rm K},(2)}_{\uparrow}&=&
%\red{\frac{1}{2}}
\left\langle\psi_{A}^{{\rm K}}\right|
\!H^{{\rm K},(2)}\!
\left|\psi_{A}^{{\rm K}}\right\rangle
% \\
%&& \red{+\frac{1}{2}\left\langle\psi_{{\mathbf k},B}^{K,(0)}\right|H^{K,(2)}
%\left|\psi_{{\mathbf k},B}^{K,(0)}\right\rangle} \nonumber\\
\!\pm\!\nu{\rm Re}\!\left[ \langle \psi_{A}^{{\rm K}}|
H^{{\rm K},(2)}
|\psi_{B}^{{\rm K}}\rangle e^{i\theta} \right]
\nonumber \\
E^{{\rm K},(2)}_{\downarrow}&=&-E^{{\rm K},(2)}_{\uparrow},
\label{eq:second-order energy correction[2-1]}
\end{eqnarray}
%%%%%%%%%%%%
%
%
where the upper (lower) sign applies to the energy shift of the conduction band
bottom (valence band top)
and
$\langle\psi_{A}^{{\rm K}}|H^{{\rm K},(2)}
|\psi_{A}^{{\rm K}}\rangle
\!=\!\langle\psi_{B}^{{\rm K}}|H^{{\rm K},(2)}
|\psi_{B}^{{\rm K}}\rangle$ is used.
Then by comparing $E^{{\rm K},(2)}_{\uparrow(\downarrow)}$
%Eqs.~(\ref{eq:second-order energy correction[2]})
%and $E^{{\rm K},(2)}_{\downarrow}$
%(\ref{eq:second-order energy correction[2-1]})
with Fig.~\ref{fig:schematic level diagram},
one finds
%
%
%%%%%%%%%%%%
\begin{equation}
\frac{\delta_{\rm K}}{R}=\left\langle\psi_{A}^{\rm K}\right|\!H^{\rm K,(2)}\!\left|\psi_{B}^{\rm K}\right\rangle, \,\,\,
\frac{\delta'_{\rm K}}{R}=\left\langle\psi_{A}^{\rm K}\right|\!H^{{\rm K},(2)}\!\left|\psi_{A}^{\rm K}\right\rangle.
\label{eq:delta delta'}
\end{equation}
%%%%%%%%%%%%
%
%
Note that $\delta_{\rm K}$ and $\delta'_{\rm K}$ are related to
pseudospin-flipping and pseudospin-conserving processes,
respectively.

%
%
%%%%%%%%%%%%%%%
\begin{figure}[b!]
\centerline{\includegraphics[width=8cm]{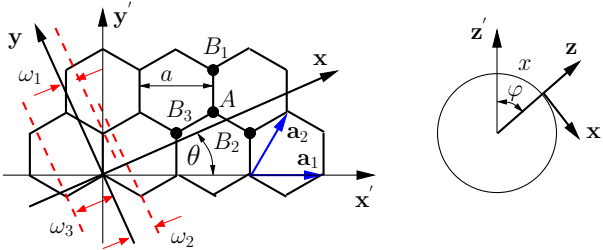}}
\caption{(Color online) Two-dimensional honeycomb lattice structure.
${\bf x} ({\bf y})$ is the coordinate around (along) the CNT with
chiral vector $n\,{\mathbf a_1}\!+\!m\,{\mathbf a_2}\!\equiv\!(n,m)$
and chiral angle $\theta$.
$\omega_{j}\,(j\!=\!1,2,3)$,
the length
between $y$ axis passing $A$ atom and its parallel (red dashed) line
is related with $\xi_{j}$ by
$\xi_{j}\!\approx\!\omega_j/(2R)$ [Eq.~(\ref{eq:curvature-H-pi-sigma})].
The coordinates for the CNT is
illustrated on the right.
Here, $x\!=\!\varphi R$.
}
\label{fig:coordinate}
\end{figure}
%%%%%%%%%%%%%%
%

%\section{Tight-binding model}
%\label{sec:TB}
%
%
To evaluate Eq.~(\ref{eq:delta delta'}),
one needs explicit expressions for $H_{\rm so}$, $H_{\rm c}$,
and $H^{\rm K,(0)}$.
$H_{\rm so}$ is given by $\lambda_{\rm so}\!\sum_r \mathbf{L}_r\!\cdot\!\mathbf{S}_r$~\cite{Min06PRB},
where $\mathbf{L}_r$ and $\mathbf{S}_r$ are respectively the atomic orbital
and spin angular momentum of an electron at a carbon atom $r$.
The tight-binding Hamiltonian of the $H_{\rm so}$ can be written~\cite{Hernando06PRB} as
%
%
%\begin{eqnarray}
%\label{eq:atomic-spin-orbit-H[1]}
%\begin{split}
$H_{\rm so}={(\lambda_{\rm so}/2)}
\sum_{ r={\bf r}_{A/B}}(c_{r-}^{z\dagger}c_{r+}^x-c_{r+}^{z\dagger}c_{r-}^x
+ic_{r+}^{z\dagger}c_{r-}^y
+ic_{r-}^{z\dagger}c_{r+}^y
+ic_{r+}^{y\dagger}c_{r+}^x-ic_{r-}^{y\dagger}c_{r-}^x)+{\rm H.c.}$,
%\end{split}
%\end{eqnarray}
%
%
where $c_{r+(\!-\!)}^{x}$, $c_{r+(\!-\!)}^{y}$, and
$c_{r+(\!-\!)}^{z}$ denote the annihilation operators for
$|p_x^r\rangle\chi_{+(\!-\!)}$, $|p_y^r\rangle\chi_{+(\!-\!)}$, and $|p_z^r\rangle\chi_{+(\!-\!)}$.
Here $\chi_{+(\!-\!)}$ denotes the eigenspinor of $\sigma_z$
($+/-$ for outward/inward).
%which is related with $\chi_{\uparrow (\downarrow)}$ by
%
%
%\begin{eqnarray}
%\label{eq:transform}
%\chi_{+}\!=\!{1\over\sqrt{2}}\left(\chi_{\uparrow}+e^{i\varphi}\chi_{\downarrow}\right),\,\,\,\,\,\,
%\chi_{-}\!=\!{i\over\sqrt{2}}\left(-\chi_{\uparrow}+e^{i\varphi}\chi_{\downarrow}\right),
%\end{eqnarray}
%
%
%where $\varphi$ is the azimuthal angle $x/R$ around the CNT axis (Fig.~\ref{fig:coordinate}).
For later convenience,
%the $y$-axis is chosen as the spin quantization axis
%and then with the help of the Eq.~(\ref{eq:transform}),
%$H_{\rm so}$ becomes}
%we express $H_{\rm so}$
%in term of $\chi_{\uparrow\,(\downarrow)}$}%using Eq.~(\ref{eq:transform}})
we express $\chi_{+(-)}$ in term of $\chi_{\uparrow (\downarrow)}$
to obtain a expression for $H_{\rm so}$,
%When the $y$-axis is chosen as the spin quantization axis,
%it becomes
%
%
%%%%%%%%%%%%%%
\begin{eqnarray}
%\begin{split}
\label{eq:atomic-spin-orbit-H[2]}
H_{\rm so}&=&\frac{\lambda_{\rm so} }{2}
\sum_{r={\bf r}_{A/B}}\left[
i\left(c_{r\downarrow}^{z\dagger}c_{r\downarrow}^x
-c_{r\uparrow}^{z\dagger}c_{r\uparrow}^x\right)\right.\nonumber\\
&+&\left(e^{-i\varphi}c_{r\uparrow}^{z\dagger}c_{r\downarrow}^y\right.
-\left.e^{i\varphi}c_{r\downarrow}^{z\dagger}c_{r\uparrow}^y\right) \nonumber\\
&+&i\!\left.\left(e^{-i\varphi}c_{r\uparrow}^{y\dagger}c_{r\downarrow}^x
+e^{i\varphi}c_{r\downarrow}^{y\dagger}c_{r\uparrow}^x\right)
\right]
+{\rm H.c.}.
%
%+&\left. e^{-i\varphi}\left(c_{r\uparrow}^{z\dagger}c_{r\downarrow}^y
%+i c_{r\uparrow}^{y\dagger}c_{r\downarrow}^x\right)
%-e^{i\varphi}\left( c_{r\downarrow}^{z\dagger}c_{r\uparrow}^y
%-i c_{r\downarrow}^{y\dagger}c_{r\uparrow}^x\right)
%%\right)\\
%%+&\left.i\left(e^{-i\varphi}
%%+e^{i\varphi}
%\right]
%+{\rm H.c.},
%\end{split}
\end{eqnarray}
%%%%%%%%%
%
%
%where $c^{s,x,y,z}_{r\alpha}$ denote the annihilation operator
%for the electron in the orbital $s$, $p_x$, $p_y$, $p_z$ at the carbon atom $r$
%with the real spin $\sigma_y\!=\!\alpha$ ($\alpha\!=\uparrow,\downarrow$).
%Here $\varphi$ is the azimuthal angle $x/R$ around the CNT axis.
%and those terms in Eq.~(\ref{eq:atomic-spin-orbit-H}) with the factors $e^{\pm i \varphi}$
%do not contribute to the effective SOC [Eq.~(\ref{eq:effective SOC Hamiltonian})]
%since their effects vanish upon the integration over the angle $\varphi$~\cite{Hernando06PRB}.
%Hence they will be ignored below.
%
%Note that $H_{\rm so}$ conserves the pseudospin. The real spin, on the other hand,
%is not strictly conserved, since the last four terms induce the real spin flip.
%Those terms, however, do not
%Then the first two terms are the only relevant terms.
%Note that they induce inter-band transition between $\pi$ and $\sigma$ bands (Fig.~\ref{fig:transition process}).

For the curvature Hamiltonian $H_{\rm c}$,
we retain only the leading order term in the expansion in terms of $a/R$.
Up to the first order in $a/R$,
$H_{\rm c}$ reduces to $H_{\rm c}^{\pi\sigma}$,
\begin{eqnarray}
\label{eq:curvature-H-pi-sigma}
H_{\rm c}^{\pi\sigma}\!&=&\!\sum_{{\mathbf r}_A}\!\sum_{j=1}^{3}\!\sum_{\alpha=\!\uparrow,\downarrow}
\!\left[
S_j\left(c_{{\mathbf r}_A\alpha}^{z\dagger}c_{B_j\alpha}^s\!
\!+c_{{\mathbf r}_A\alpha}^{s\dagger}c_{B_j\alpha}^{z}\right)\right.
\! \nonumber \\
&+& \!X_j\!\left(c_{{\mathbf r}_A\alpha}^{z\dagger}c_{B_j\alpha}^x
\!-c_{{\mathbf r}_A\alpha}^{x\dagger}c_{B_j\alpha}^z\right) \\
&+& \!Y_j\!\left.\left(c_{{\mathbf r}_A\alpha}^{z\dagger}c_{B_j\alpha}^y
\!-c_{{\mathbf r}_A\alpha}^{y\dagger}c_{B_j\alpha}^z\right)\right]\!+{\rm H.c.},
\nonumber
%+&\xi_2\left\{
%U_4\left(c_{A\tau}^{z\dagger}c_{B_2\tau}^s+c_{A\tau}^{s\dagger}c_{B_2\tau}^z\right)
%+U_5\left(c_{A\tau}^{z\dagger}c_{B_2\tau}^x\!-\!c_{A\tau}^{x\dagger}c_{B_2\tau}^z\right)\!
%+U_6\left(c_{A\tau}^{z\dagger}c_{B_2\tau}^y\!-\!c_{A\tau}^{y\dagger}c_{B_2\tau}^z\right)\right\} \\
%+&\xi_3\left.\left\{
%U_7\left(c_{A\tau}^{z\dagger}c_{B_3\tau}^s+c_{A\tau}^{s\dagger}c_{B_3\tau}^z\right)
%+U_8\left(c_{A\tau}^{z\dagger}c_{B_3\tau}^x\!-\!c_{A\tau}^{x\dagger}c_{B_3\tau}^z\right)\!
%+U_9\left(c_{A\tau}^{z\dagger}c_{B_3\tau}^y\!-\!c_{A\tau}^{y\dagger}c_{B_3\tau}^z\right)\right\}\right]\!\\
%+&\,{\rm H.c},
\end{eqnarray}
where
${\mathbf r}_A$ is a lattice site in the sublattice $A$ and
its three nearest neighbor sites in the sublattice $B$
are represented by $B_j$ ($j\!=\!1,2,3$) (Fig.~\ref{fig:coordinate}).
Here $S_j$, $X_j$, $Y_j$ are proportional to $a/R$ and denote the curvature-induced coupling strengths of
$s$, $p_x$, $p_y$ orbitals with a nearest neighbor $p_z$ orbital.
Their precise expressions that
can be determined purely from geometric considerations,
are given by
%\begin{equation}
%\label{eq:parameters[1]}
$S_j\!=\!\xi_j\tilde{S}_j$,
$X_j\!=\!\xi_j\tilde{X}_j$, and
$Y_j\!=\!\xi_j\tilde{Y}_j$
%\end{equation}
%
%
with
%\begin{equation}
%\begin{split}
%\label{eq:parameters[2]}
$\xi_1\approx{a/(2\sqrt{3}R)}\sin\theta$, %\,\,\,\,\,\,
$\xi_2\approx{a/(2\sqrt{3}R)}\sin\left({\pi/3}-\theta\right)$, and
$\xi_3\approx{a/(2\sqrt{3}R)}\sin\left({\pi/3}+\theta\right)$(Fig.~\ref{fig:coordinate}).
%\end{split}
%\end{equation}
%
%
Here
\begin{eqnarray}
%\begin{split}
\label{eq:parameters[3]}
\tilde{S}_1&=&V_{sp}^{\sigma}\sin\theta,\nonumber\\
\tilde{S}_2&=&V_{sp}^{\sigma}\cos\left({\pi\over6}+\theta\right), \nonumber\\
\tilde{S}_3&=&V_{sp}^{\sigma}\cos\left({\pi\over6}\!-\!\theta\right), \nonumber\\
\tilde{X}_1&=&-V_{pp}^{\sigma}\sin^2\theta\!-\!V_{pp}^{\pi}-V_{pp}^{\pi}\cos^2\theta\!,\nonumber\\
\tilde{X}_2&=&-V_{pp}^{\sigma}\sin^2\left({\pi\over3}-\theta\right)-V_{pp}^{\pi}
-V_{pp}^{\pi}\cos^2\left({\pi\over3}-\theta\right), \nonumber\\
\tilde{X}_3&=&V_{pp}^{\sigma}\sin^2\left({\pi\over6}-\theta\right)+V_{pp}^{\pi}
+V_{pp}^{\pi}\cos^2\left({\pi\over6}-\theta\right), \nonumber\\
\tilde{Y}_1&=&\sin(2\theta){V_{pp}^{\pi}\!-\!V_{pp}^{\sigma}\over2},\nonumber\\
\tilde{Y}_2&=&\sin\left(2\theta\!-\!{2\pi\over3}\right){V_{pp}^{\pi}\!-\!V_{pp}^{\sigma}\over2}, \nonumber\\
\tilde{Y}_3&=&\sin\left(2\theta\!-\!{\pi\over3}\right){V_{pp}^{\pi}\!-\!V_{pp}^{\sigma}\over2}.
%\end{split}
\end{eqnarray}
%
%
%
%Note that $H_{\rm c}^{\pi\sigma}$ always induces the pseudospin flip and
%the inter-band transitions (Fig.~\ref{fig:transition process}). It conserves only the real spin.

Lastly, for the factor $H^{\rm K,(0)}$,
we use the Slater-Koster parametrization~\cite{Slater54PR}
for nearest-neighbor hopping.
In $\sigma$ band calculation, $s$, $p_x$, and $p_y$ orbitals are used as basis.

%in order to calculate the $\pi$ (Eq.~\ref{eq:DiracHamiltonian})
%and $\sigma$ bands
%using ($s$, $p_x$, $p_y$, $p_z$) orbitals as
%orthogonal basis

%It is evident that it does not induces the real spin flip nor the inter-band transition.
%For the pseudospin, however, it may or may not induce a flip (Fig.~\ref{fig:transition process}),
%since states localized within one particular sublattice are not eigenstates of $H^{\rm K,(0)}$.
%Amplitudes of the pseudospin flip and nonflip are similar in magnitude.

%
%
%%%%%%%%%%%%%%%%%%%%%%%%%%%%%
%%%%%%%
%
%
%

%\subsection{Expressions of $\delta$ and $\delta'$}
Combined effects of the three factors
$H_{\rm so}$, ${\cal P}^{\rm K}/(E^{(0)}-H^{\rm K,(0)})$, $H_{\rm c}$
%can be easily read off.
are illustrated in Fig.~\ref{fig:transition process}.
The real spin dependence arises solely from $H_{\rm so}$,
which generates the factor $\sigma_y$~\cite{comment-sigma_y}.
For the pseudospin,
the combined effect of $H_{\rm so}$ and $H_{\rm c}$ is to flip the pseudospin.
When they are combined with the pseudospin conserving part of ${\cal P}^{\rm K}/(E^{(0)}\!-\!H^{\rm K,(0)})$,
one obtains the pseudospin flipping process [Eq.~(\ref{eq:delta delta'})]
determining $\delta_{\rm K}$.
In addition, ${\cal P}^{\rm K}/(E^{(0)}\!-\!H^{\rm K,(0)})$
contains the pseudospin flipping part, which is natural
%The existence of the pseudospin flipping part of ${\cal P}^{\rm K}/(E^{(0)}\!-\!H^{\rm K,(0)})$ is natural
since states localized in one particular sublattice are not eigenstates of $H^{\rm K,(0)}$.
%\comment{Jae-Seung, is the pseudospin flipping part of ${\cal P}^{\rm K}/(E^{(0)}\!-\!H^{\rm K,(0)})$ comparable
%to the pseudospin conserving part in magnitude?}
When the pseudospin flipping part of ${\cal P}^{\rm K}/(E^{(0)}-H^{\rm K,(0)})$
is combined with $H_{\rm so}$ and $H_{\rm c}$,
one obtains the pseudospin conserving process [Eq.~(\ref{eq:delta delta'})]
determining $\delta'_{\rm K}$.

The signs of $\delta_{\rm K} e^{i\theta}$ and $\delta'_{\rm K}/\cos3\theta$ are negative.
We find $|(\delta'_{\rm K}/\cos 3\theta)/\delta_{\rm K}|=4.5$
for tight-binding parameters in Ref.~\cite{Mintmire95Carbon}.
%\comment{Jae-Seung, please provide numerical values of this ratio}
Thus $\delta'_{\rm K}$ is of the same order as
$\delta_{\rm K}$~\cite{parameters}, which is understandable since pseudospin flipping terms
in $E^{(0)}\!-\!H^{\rm{K},(0)}$ (with amplitudes $V_{pp}^{\sigma}$, $V_{sp}^{\sigma}$)
are comparable in magnitude to pseudospin conserving terms
(with amplitudes $E^{(0)}\!-\!\varepsilon_{s (p)}$).
%can be larger or smaller than $\delta$ depending on $\theta$.

%
%By combining the expressions for the three factors in Eq.~(\ref{eq:H^K(2)}),
%one finds
%\begin{eqnarray}
%{\delta\over L}\!&=&
%%\!\!\left\langle\psi_{A}^{\rm K}\right|\!H^{\rm K,(2)}\!\left|\psi_{B}^{\rm K}\right\rangle
%%\!\=
%\!\frac{\lambda_{\rm so}(\varepsilon_{s}\!-\!\varepsilon_{p}) \pi a (V_{pp}^{\pi}\!+\!V_{pp}^{\sigma})e^{-i\theta}}
%{6\sqrt{3}{V_{sp}^{\sigma}}^2L},
%\label{eq:delta} \\
%{\delta'\over L}\!\!&=&
%%\!\!\left\langle\psi_{A}^{\rm K}\right|\!H^{{\rm K},(2)}\!\left|\psi_{A}^{\rm K}\right\rangle
%%\!=
%\!\frac{\lambda_{\rm so}\pi a V_{pp}^{\pi}\cos3\theta}{\sqrt{3}(V_{pp}^{\sigma}-V_{pp}^{\pi})L},
%\label{eq:delta'}
%\end{eqnarray}
%where

%Physical meanings of $\delta\sigma_y$ and $\delta'\sigma_y$ are different.
%While $\delta\sigma_y$ is related to the spin-dependent
%topological phase induced by the curvature~\cite{Ando00JPSJ,Hernando06PRB,Kuemmeth08Nature},
%$\delta'\sigma_y$ can be interpreted as the coupling between
%the Bloch momentum along the circumference and the spin
%\red{In agreement with this interpretation,
%we find}
%\blue{on the ground [Eq.~(\ref{eq:delta'})]} that $|\delta'|$ is maximal for zigzag CNTs ($\theta\!=\!0$),
%for which ${\mathbf K}$ is parallel with the $x$ axis,
%and vanishes for armchair CNTs ($\theta\!=\!\pi/6$),
%for which ${\mathbf K}$ is perpendicular to the $x$ axis\blue{~\cite{K'-delta dependence}}.

\section{Behavior in a magnetic field}
\label{sec:Behavior in a magnetic field}
\begin{figure}[b!]
\centerline{\includegraphics[width=8.5cm]{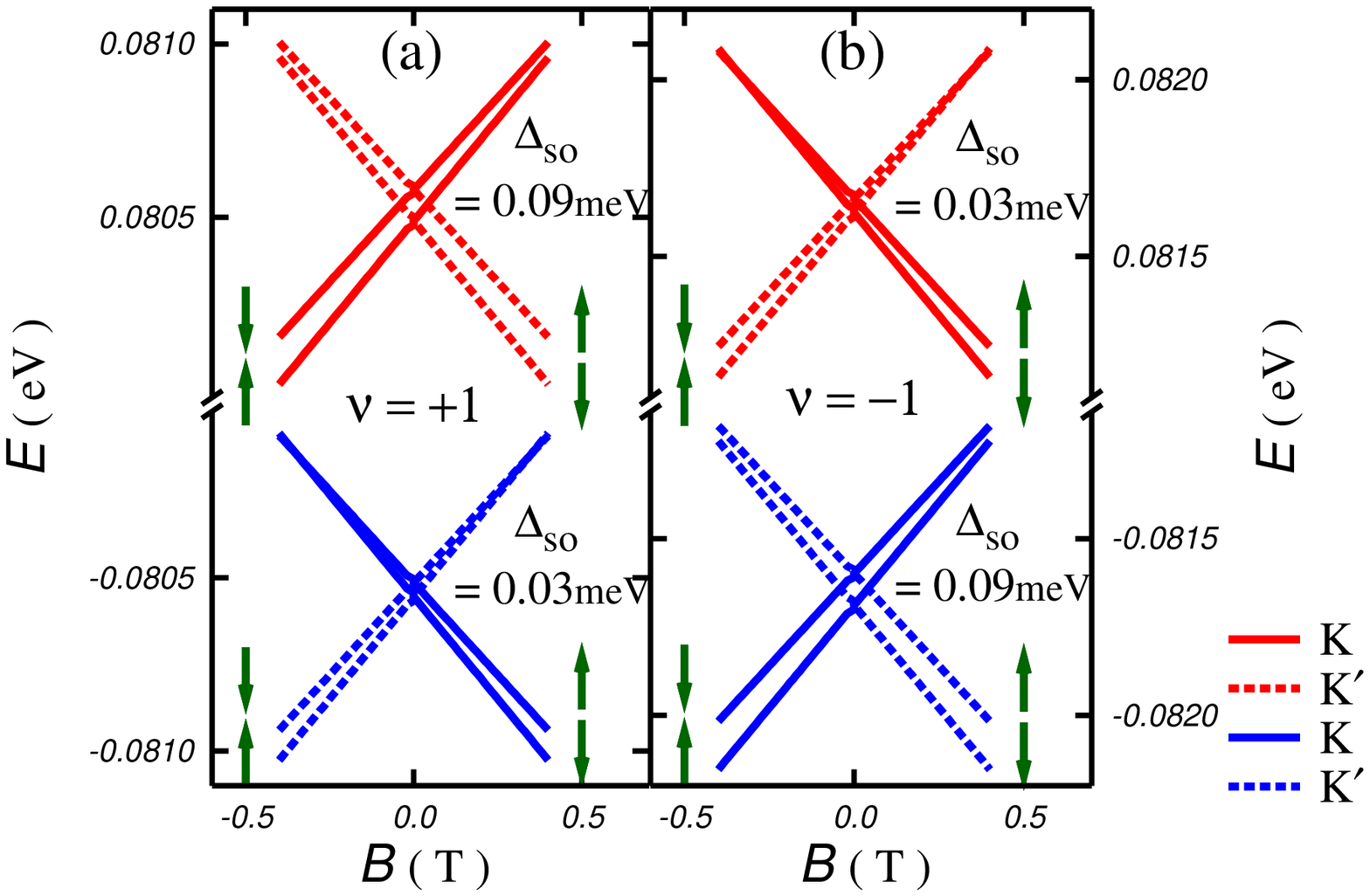}}
\centerline{\includegraphics[width=8.5cm]{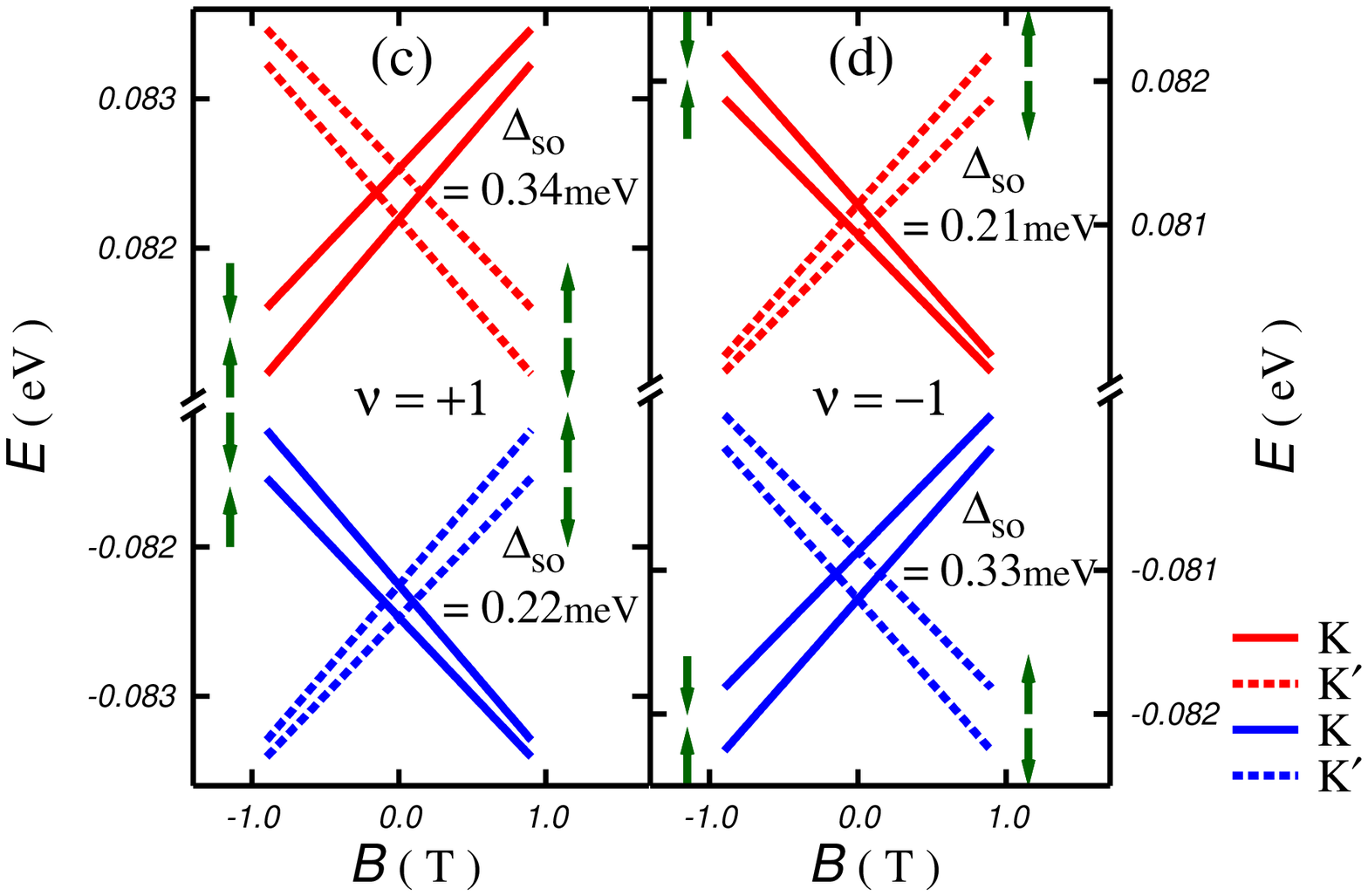}}
\caption{(Color online)
%\comment{Please define $\Delta_{\rm so}$
%somewhere and polish expressions in this caption.}
Calculated energy spectrum of the conduction band bottom (red, $E>0$)
and valence band top (blue, $E<0$) near
K (solid lines) and K$'$ (dashed lines) points
in semiconducting CNTs with $R\!\approx\!2.5\,$nm
as a function of magnetic field $B$
parallel to the CNT axis. The chiral vectors
for each CNT
are (a) (38,34), (b) (39,34), (c) (61,0), and (d) (62,0),
respectively.
Arrows (green) show spin direction along the CNT axis
%represent eigenspinor $\chi_{\uparrow/\downarrow}$
and $\Delta_{\rm so}$ denotes the zero-field splitting.
Assuming $k_y\!=\!0$,
the energy $E$ including the SOC,
%[Eqs.~(\ref{eq:second-order energy correction[2]})
%and~(\ref{eq:second-order energy correction[3]})],
the Aharonov-Bohm flux~\cite{Ajiki93JPSJ} %induced by $B$,
$\phi_{\rm AB}\!=\!{B\pi R^2}$,
and the Zeeman coupling effects
is,
$E\!=\!\pm\hbar v_{F} \sqrt{(k_x\!+\!{(1/R)}{(\phi_{\rm AB}/ \phi_0)})^2}
+E_{\uparrow(\downarrow)}^{{\rm K(K')},(2)}\!+\!{(g/2)}\mu_{\rm B}\tau_{\|} B$,
with upper (lower) sign applying to the conduction (valence) band.
$\phi_0\!=\!{hc/|e|}$,
$\tau_{\|}\!=\!+1(-1)$ for $\chi_{\uparrow(\downarrow)}$,
$v_{F}\!=\!-aV_{pp}^{\pi}{\sqrt{3}/2}$,
and $g\!=\!2$~\cite{Kuemmeth08Nature}.
%is the gyromagnetic ratio,
%Here $k_{x}\!=\!2\pi/(3L)\nu$ is near $K$ (solid lines) and
%$k_{x}\!=\!-2\pi/(3L)\nu$ near $K'$ (dashed lines).
For estimation of $E_{\uparrow(\downarrow)}^{{\rm K(K')},(2)}$, we
use tight-binding parameters in Ref.~\cite{Mintmire95Carbon};
$V_{ss}^{\sigma}\!=\!-4.76\,$eV, $V_{sp}^{\sigma}\!=\!4.33\,$eV,
$V_{pp}^{\sigma}\!=\!4.37\,$eV, $V_{pp}^{\pi}\!=\!-2.77\,$eV,
$\varepsilon_s\!=\!-6.0\,$eV, and
$\varepsilon_p\!=\!0$~\cite{parameters}. }
%$V_{sp}^{\sigma}\!=\!-5.58\,$eV, $V_{pp}^{\sigma}\!=\!-5.037\,$eV, $V_{pp}^{\pi}\!=\!-3.033\,$eV,
%and $V_{ss}^{\sigma}\!=\!-6.769\,$eV.
%The overlap between nearest neighbor orbitals with $S_{ss}^{\sigma}\!=\!0.212$,
%$S_{sp}^{\sigma}\!=\!0.102$, $S_{pp}^{\sigma}\!=\!0.146$,
%and $S_{pp}^{\pi}\!=\!0.129$, are also considered in calculating eigenstates and eigenenergy
%of $\sigma$ bands~\cite{parameters}}.
\label{fig:SOC}
\end{figure}

Next we examine further implications of our result in view of the experiment~\cite{Kuemmeth08Nature},
where the conduction band bottom and valence band top positions of semiconducting CNTs ($\nu\!=\!\pm 1$)
are measured as a function of the magnetic field $B$ parallel to the CNT axis.
We find that the $\theta$ dependence [Eq.~(\ref{eq:delta'})] of $\delta'_{\rm K}$ has interesting implications.
When $\cos 3\theta$ is sufficiently close to $0$ (close to armchair-type),
$|\delta'_{\rm K}|$ is smaller than $|\delta_{\rm K} e^{i\theta}|$.
The prediction of our theory in this situation is shown in Figs.~\ref{fig:SOC}(a) and (b).
Note that the spin splitting of both the conduction and valence bands becomes smaller as the energy $E$ increases.
%which agrees with the experimental result~\cite{Kuemmeth08Nature}.
On the other hand, when $\cos 3\theta$ is sufficiently close to $1$ (close to zigzag-type),
$|\delta'_{\rm K}|$ is larger than $|\delta_{\rm K} e^{i\theta}|$.
In this situation [Figs.~\ref{fig:SOC}(c) and (d)],
the energy dependence of either valence or conduction band is inverted;
For $\nu\!=\!+1 (-1)$, the spin splitting of the valence (conduction) band becomes {\it larger}
as the the energy increases.

Combined with the electron-prevailing
[Figs.~\ref{fig:SOC}(a) and (c) for $\nu\!=\!+1$]
vs. hole-prevailing [Figs.~\ref{fig:SOC}(b) and (d) for $\nu\!=\!-1$]
asymmetries
in the zero-field splitting,
one then finds that there exist four distinct patterns of $E$ vs. $B$ diagram,
which is the second main result of this paper.
Among these 4 patterns, only the pattern in Fig.~\ref{fig:SOC}(a) is observed in
the experiment~\cite{Kuemmeth08Nature}, which measured two CNT samples.
We propose further experiments to test the existence of the other three patterns.
%
%
%%%%%%%%%%%%%%%%%%%%%%%%%%%%%%%
%%%%%%%%%%%%%%%%%%%%%%%%%
%
%

Here we remark that although Eqs.~(\ref{eq:effective SOC Hamiltonian}), (\ref{eq:delta}),
(\ref{eq:delta'})
are demonstrated so far for semiconducting CNTs,
they hold for metallic CNTs ($\nu\!=\!0$) as well.
For armchair CNTs with $\cos3\theta\!=\!0$, $\delta'_{\rm K}$ becomes zero
and the spin splitting is determined purely by $\delta_{\rm K}$.
%and one finds the symmetric SOC.
For metallic but non-armchair CNTs,
finding implications of Eq.~(\ref{eq:effective SOC Hamiltonian})
is somewhat technical
since the curvature-induced minigap appears near the Fermi level~\cite{Kane97PRL}.
%we calculate the curvature-induced gap considering
%both misalignment between two nearest $p_z$ orbitals~\cite{curvature-H-pi-pi} and $H_{\rm c}^{\pi\sigma}$
%as perturbations.
Our calculation for $(37,34) (\cos3\theta\!\approx\!0)$
and $(60,0) (\cos3\theta\!=\!1)$ CNTs including the minigap effect indicates that
they show behaviors similar to Fig.~\ref{fig:SOC}(b) and (d), respectively.
Thus nominally metallic CNTs exhibit spin splitting patterns of $\nu\!=\!-1$ CNTs.
%``metallic'' nonarmchair
%the $B$-dependence
%and the zero-field spin splitting
%for lowest conduction and highest valence bands
%are similar to those of CNTs with $\nu\!=\!-1$
%[Fig.~\ref{fig:SOC}(b) and (d)]
%since the curvature effects lowers (raises)
%energy of bonding (antibonding) degenerate state~\cite{Kane97PRL}.
%
%In metallic armchair CNTs where no gap appears near the Fermi level,
%which is also confirmed
%%\comment{????}
%in our calculation,
%we obtain the symmetric SOC ($\delta'\!=\!0$).

\section{Discussion and summary}
\label{sec:Discussion and summary}
Lastly we discuss briefly the effective SOC in a curved graphene~\cite{Meyer07Nature}.
Unlike CNTs, there can be both convex-shaped and concave-shaped curvatures in a graphene.
We first address the convex-shaped curvatures.
When the local structure of a curved graphene has two principal curvatures,
$1/R_1$ and $1/R_2$ with the corresponding binormal unit vectors
${\bf n}_1$ and ${\bf n}_2$, each principal curvature
$1/R_i$ $(i\!=\!1,2)$ generates the effective SOC,
Eq.~(\ref{eq:effective SOC Hamiltonian}),
with $\sigma_y$ replaced by ${\bm \sigma}\cdot{{\bf n}_i}$ and $R$ by $R_i$.
The corresponding $\delta_{i}$ and $\delta'_{i}$ values are given
by Eqs.~(\ref{eq:delta}) and~(\ref{eq:delta'})
with $\theta$ replaced by $\theta_i$,
where $\theta_i$ is the chiral angle with respect to ${\bf n}_i$.
Thus the diagonal term
of the effective SOC is again comparable in magnitude to the off-diagonal term.
For the concave-shaped curvatures, we find that the two types of the SOC become
$-\delta_{i}$ and $-\delta'_{i}$ with $\theta_i$, respectively.
We expect that this result may be relevant
for the estimation of the spin relaxation length in graphenes~\cite{Hernando08Condmat}
and may provide insights into unexplained experimental data
in graphene-based spintronic systems~\cite{Han09PRL}.
We also remark that the effective SOC in a graphene may be
spatially inhomogeneous since the local curvature of the nanometer-scale
corrugations~\cite{Meyer07Nature} is not homogeneous,
whose implications go beyond the scope of this paper.
%\comment{Please check this paragraph}

In summary, we have demonstrated that the interplay of the atomic SOC and the curvature generates two types of
the effective SOC in a CNT, one of which was not recognized before.
Combined effects of the two types of the SOC in CNTs explain
recently observed electron-hole asymmetric spin splitting~\cite{Kuemmeth08Nature}
and generates four qualitatively different types of energy level dependence
on the parallel magnetic field.
Our result may have interesting implications for graphenes as well.

{\it Note added.--} While we were preparing our manuscript, we became aware
of a related paper~\cite{Chico09PRB}.
However the effective Hamiltonian [Eq.~(\ref{eq:effective SOC Hamiltonian})] for the SOC
and the four distinct types of the magnetic field dependence (Fig.~\ref{fig:SOC})
are not reported in the work.

%%%%%%%%%%%%%%%%%%%%%%%%%%%%%%%%%%%%%%%%%%%%%%%%%%%%%%%%%%%%%%%%%%%%%%%%%%%
\begin{acknowledgements}
We appreciate Philp Kim for his comment for the curved graphenes.
We acknowledge the hospitality of
Hyunsoo Yang
and Young Jun Shin
at National University of Singapore, where parts of
this work were performed.
We thank Seung-Hoon Jhi, Woojoo Sim, Seon-Myeong Choi and Dong-Keun Ki for
helpful conversations.
This work was supported by the KOSEF (Basic Research Program No. R01-2007-000-20281-0)
and BK21.
\end{acknowledgements}
%%%%%%%%%%%%%%%%%%%%%%%%%%%%%%%%%%%%%%%%%%%%%%%%%%%%%%%%%%%%%%%%%%%%%%%%%%%%%%%%%%%%%%%%%

%\comment{Provide the value of $\lambda_{\rm so}$ somewhere}
%\comment{It seems that some references are not cited at all. Please remove them.}
%\comment{Please order references in the order of appearance.}

%%%%%%%%%%%%%%%%%%%%%%%%%%%%%%%%%%%%%%%%%%%
%\appendix
%%%%%%%%%%%%%%%%%%%%%%%%%%%%%%%%%%%%%%%%%%%%%%
%\section{curvature Hamiltonian}
%\label{curvature coefficient}

%%%%%%%%%%%%%%%%%%%%%%%%%%%%%%%%%%%%%%%%%%%%%%%%%%%

%%%%%%%%%%%%%%%%%%%%%%%%%%%%%%%%%%%%%%%%%%%%%%%%%%


\begin{thebibliography}{99}
\bibitem{comment-delta}
The corresponding expression in Ref.~\cite{Hernando06PRB}
is slightly different from Eq.~(\ref{eq:delta})
since
the $\sigma$ band is treated in different ways.
%However their numerical values are similar.
When a few minor mistakes in Ref.~\cite{Hernando06PRB} are corrected,
the two expressions result in similar numerical values.
\bibitem{Second order Hamiltonian}
$H_{\rm c}[\mathcal{P}^{\rm K}/(E^{(0)}\!-\!H^{{\rm K},(0)})]H_{\rm c}$
is spin-independent and thus ignored [see Eq.~(\ref{eq:curvature-H-pi-sigma})].
\bibitem{comment-sigma_y}
The last four terms of $H_{\rm so}$ in Eq.~(\ref{eq:atomic-spin-orbit-H[2]}),
which do not commute with $\sigma_y$,
do not contribute to the effective SOC near the Fermi energy
due to the factor $e^{\pm i\varphi}$~\cite{Hernando06PRB}.
\end{thebibliography}
\end{document}